\documentclass[10pt]{iopart}

\usepackage{graphicx}
\usepackage{xcolor}
\begin{document}

\title[Quantum beats in fluorescence of helium]{Time-resolved quantum beats in the fluorescence of helium resonantly excited by XUV radiation}

\author{A.~C.~LaForge$^{1,*}$, A.~Benediktovitch$^{2,*}$, V.~Sukharnikov$^{2,*}$, {\v{S}}.~Kru{\v{s}}i{\v{c}}$^3$, M.~{\v{Z}}itnik$^3$, M.~Debatin$^4$, R.~W.~Falcone$^5$,  J. D.~Asmussen$^6$, M.~Mudrich$^6$, R.~Michiels$^7$, F.~Stienkemeier$^7$,  L.~Badano$^8$, C.~Callegari$^8$, M.~Di~Fraia$^8$,  M.~Ferianis$^{8,9}$, L.~Giannessi$^{8,9}$, O.~Plekan$^8$, K.~C.~Prince$^8$, C.~Spezzani$^8$, N.~Rohringer$^{2,10}$, and N.~Berrah$^1$ }

\address{$^1$ Department of Physics, University of Connecticut, Storrs, Connecticut, 06269, USA}
\address{$^2$ Deutsches Elektronen-Synchrotron (DESY), 22607 Hamburg, Germany}
\address{$^3$ Jo{\v{z}}ef Stefan Institute, Jamova cesta 39, SI-1001 Ljubljana, Slovenia}
\address{$^4$ Experimentalphysik I, Universität Kassel, D-34132 Kassel, Germany}
\address{$^5$ Department of Physics, UC Berkeley, Berkeley, California 94720, USA}
\address{$^6$ Department of Physics and Astronomy, Aarhus University, 8000 Aarhus C, Denmark}
\address{$^7$ Physikalisches Institut, Universit{\"a}t Freiburg, 79104 Freiburg, Germany}
\address{$^8$ Elettra-Sincrotrone Trieste, 34149 Basovizza, Trieste, Italy}
\address{$^9$ Nazionale di Fisica Nucleare - Laboratori Nazionali di Frascati,  Via E. Fermi 40, 00044 Frascati, Roma, Italy}
\address{$^{10}$ Department of Physics, Universität Hamburg, Hamburg 22761, Germany} 
\address{$^*$ These authors contributed equally to this work.}

\ead{aaron.laforge@uconn.edu,andrei.benediktovitch@cfel.de}	

\vspace{10pt}

\begin{abstract}
We report on the observation of time-resolved quantum beats in the helium fluorescence from the transition $1s3p\,\rightarrow\,1s2s$, where the initial state is excited by XUV free electron laser radiation. The quantum beats originate from the Zeeman splitting of the magnetic substates due to an external magnetic field. We perform a systematic study of this effect and discuss the possibilities of studying this phenomenon in the XUV and X-ray regime. 
%To obtain the temporal profiles of fluorescence, a two dimensional streak camera was used. 
\end{abstract}

%\maketitle

\section{Introduction}

A system in a coherent superposition of states is a fundamental quantum-mechanical phenomenon and is at the core of numerous chemical and biological transient phenomena \cite{Zewail2000Femtochemistry}. The understanding of these phenomena is of the utmost importance and motivates the development of new experimental techniques and theoretical modeling. The essential component of the corresponding experimental techniques is the creation of the superposition of excited states which is then probed with one or several pulses after a certain delay. This approach is realized for numerous techniques such as multidimensional spectroscopy~\cite{mukamel2009}, transient absorption spectroscopy~\cite{goulielmakis2010as-coherence}, and phase-modulated coherent spectroscopy~\cite{wituschek2020phase-manipulated}, just to mention a few examples. The typical requirement is for femto- to attosecond pulse durations, dictated by the timescale of the process of interest. When the timescale of the process is long enough, instead of a second probe pulse, time-resolved fluorescence spectroscopy is a possible alternative technique. If the initial state is in a coherent superposition of excited states, the intensity of emitted fluorescence would show temporal oscillations, referred to as quantum beats.

For fluorescence from an ensemble of identical non-interacting atoms, quantum beats emerge from close lying electronic states~\cite{Aleksandrov1973,haroche1976QBreview}, e.g. from a multiplet due to Zeeman splitting of an electronic level by an external magnetic field. To illustrate the concept, consider a $V$-type level scheme \cite{scully1980quantum-beats}, which is prepared in a coherent superposition of excited states with a small energy gap, $\hbar\Delta$, between them. The frequency of the beating will be proportional to the energy gap between the excited states~\cite{haroche1976QBreview}. Due to the coherence of the emitted fluorescence, the combination of respective waves with slightly different frequencies results in the periodic modulation of the intensity of fluorescence in time. An example of the beating phenomenon is given in Fig.\,\ref{fig:2}\,(a). For the case of no external magnetic field, the fluorescence lifetime is a simple exponential decay with a time constant, $\tau$, which is given by the red line. The introduction of an external magnetic field $B_0$ causes a splitting proportional to the magnetic field strength $\hbar\Delta \sim \mu_B B_0$, producing a temporal interference in the decay, given by the green dashed line. Due to their inverse relationship, a doubling of the magnetic field reduces the period of the interference by half, given by the dash-dotted blue line.

%\textcolor{red}{ACL: This paragraph doesn't fit too well. Maybe we just drop it?} To initiate the beating, the primary requirement is the pump pulse needs to be shorter than the lifetime of the excited levels and the beating period~\cite{Aleksandrov1973}. The first proof-of-principle experimental demonstration of optical fluorescence quantum beats were performed with a shuttered spectral lamp~\cite{Aleksandrov1964exp,dodd1964exp}. Subsequent improvements in experimental techniques (e.g. impulsive electronic excitation and pulsed lasers) transformed the fluorescence quantum beating technique into a tool for high-precision optical spectroscopy~\cite{Aleksandrov1973, haroche1976QBreview, Aleksandrov_1979}. 

Since fluorescence quantum beating is a general phenomenon~\cite{scully1980quantum-beats}, it can be observed for various kinds of transitions and photon energies.  To initiate the beating, the primary requirement is that the pump pulse needs to be shorter than the lifetime of the excited levels and the beating period~\cite{Aleksandrov1973}. The first proof-of-principle experimental demonstrations of optical fluorescence quantum beats were performed with a shuttered spectral lamp~\cite{Aleksandrov1964exp,dodd1964exp}. Subsequent improvements in experimental techniques (e.g. impulsive electronic excitation and pulsed lasers) transformed the fluorescence quantum beating technique into a tool for high-precision optical spectroscopy~\cite{Aleksandrov1973, haroche1976QBreview, Aleksandrov_1979}. In the optical regime, quantum beats-related phenomena were observed for the resonant fluorescence of atoms~\cite{schenck1973time, haroche1973hyperfine} and molecules~\cite{wallenstein1974observation, chaiken1981average, bitto1990molecular}, from an ensemble of two-level atoms~\cite{ficek1987quantum, ficek1990quantum}, in quantum well systems~\cite{gobel1990quantum}, and even in the temporal profiles of superfluorescence~\cite{vrehen1977quantum, bartholdtsen1994superfluorescent, boursey1993superfluorescence}. In the XUV regime, quantum beats were used as a tool to evaluate the polarization of synchrotron radiation~\cite{hikosaka_zeeman_2020}. In the $\gamma$-ray regime, a phenomenon similar to quantum beats exists from coherently scattered radiation from nuclei M{\"o}ssbauer transitions~\cite{gerdau1986Fe57QB, gerdau_nuclear_1999, shvydko1999QB_SR}. 

In this work, we demonstrate the quantum beating phenomenon from helium atoms excited by seeded XUV free electron laser (FEL) pulses. The FEL was tuned to the resonant transition from the $1s^2$ ground state to the $1s3p$ excited state. The observed fluorescent emission occurs between the $1s3p$ and $1s2s$ excited states. Due to the coupling to an external magnetic field, the excited state is split into its magnetic substates and an interference is observed in the subsequent fluorescence intensity. Compared to previous work~\cite{hikosaka_zeeman_2020}, we have used orders of magnitude higher magnetic field strengths resulting in orders of magnitude faster beating, which can only be observed with the aid of a femtosecond streak-camera. The quantum beating effect is systematically investigated by varying the experimental parameters. We additionally discuss the prospects for quantum beating in the XUV and X-ray regime.

\section{Experimental setup}
The experiment was performed at the Low Density Matter endstation~~\cite{Lyamayev2013} of the seeded FEL FERMI, in Trieste, Italy. The FEL photon energy was set to $h\nu\,=\,23.00$\,eV, corresponding to the $1s3p$ excited state in helium, and tuned via the seed laser and the undulator gaps, with a pulse length of approximately 100\,fs full width half maximum (FWHM)~\cite{Allaria2012,Allaria2012a}. The FEL pulse energy at the sample varied from approximately 0.6\,$\mu$J to 66\,$\mu$J with a spot size of 35\,$\mu$m FWHM, and was calculated from the value measured upstream on a shot-by-shot basis by gas ionization and the nominal reflectivity of the optical elements in the beam transport system~\cite{Svetina2015}. A cold, supersonic gas jet of He atoms was produced by expansion of high pressure He gas through a pulsed, cryogenically cooled Even-Lavie nozzle. By varying the expansion conditions (backing pressure and nozzle temperature), the target density was varied in the range of 10$^{12}$-10$^{15}$ atoms/cm$^3$. The atomic beam was perpendicularly crossed by the FEL at the center of a magnetic bottle spectrometer (MBS), which measures kinetic energies of electrons from their time-of-flights, and applies an intense magnetic field~\cite{Matsuda2011}. The MBS was not directly used in this experiment, but its movable permanent magnet was used to induce the splitting of the $1s3p$ state. The FEL light was linearly polarized perpendicular to the magnetic field of the MBS. To measure the fluorescence lifetime, a streak camera (Hamamatsu Fesca-200) was mounted downstream of the interaction region on the rear portion of the endstation. Focusing optics coaxial with the FEL beam were used to collect the maximum amount of fluorescence.   

\section{Results and Discussion}
		\begin{figure}
		\centering 
		\includegraphics[width=0.65\linewidth]{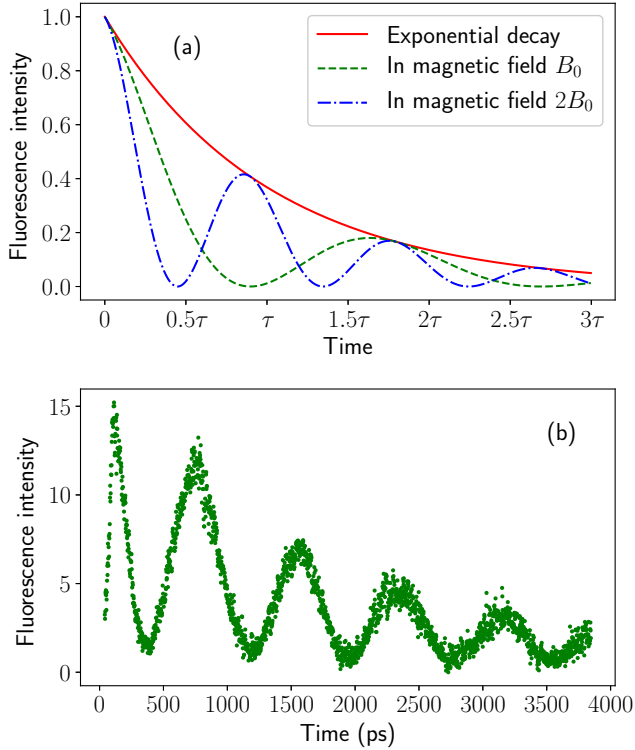}
		\caption{(a) Representative fluorescence lifetime with an exponential decay, $\tau$, for no external magnetic field (solid red line), a magnetic field $B_0$ (dashed green line), and a magnetic field $2B_0$ (dash-dotted blue line). (b) Experimentally-measured fluorescence intensity for He for the $1s3p\,\rightarrow\,1s2s$ transition. The nozzle temperature $T = 30$ K and the pulse energy $W = 66\,\mu J$ were fixed.}
		\label{fig:2}
	\end{figure} 

Fig.\,\ref{fig:2}\,(b) shows an experimentally measured temporal profile of the fluorecence intensity. To ensure the fluorescence originated from the $1s3p\,\rightarrow\,1s2s$ transition ($504\,$nm), we used a 10\,nm bandpass filter centered at 505\,nm and performed background subtraction measurements. As a cross check, we installed a polarizing filter. When changing the polarization from perpendicular to parallel with respect to the magnetic field, the fluorescence intensity vanished, indicating that the emitted fluorescence had the same polarization as the FEL excitation pulse. The maximal range of measurement of the streak camera is 500 ps while the lifetime of the $1s3p\,\rightarrow\,1s2s$ transition is $1/\Gamma\,=\,1715\,$ps. For this reason, we sequentially varied the streak camera delay by multiples of $300$ ps in order to obtain the entire fluorescence curve. For the measurement in Fig.\,\ref{fig:2}\,(b), the permanent magnet of the MBS was 16.5 mm from the interaction region. The results in (b) show beating imposed on exponential decay, which clearly indicates interference in the fluorescence emission from the Zeeman splitting of the excited state and is in agreement with calculated quantum beating shown in Fig.\,\ref{fig:2}\,(a). 

To understand the quantum beats observed in Fig.\,\ref{fig:2}\,(b), we consider an isolated helium atom in its ground state, $1s^2$, placed in a static magnetic field in the $z$-direction, $\mathbf{B} = B_0 \cdot \mathbf{e}_z$. The atom is excited by a FEL pulse propagating in the $y$-direction, which is linearly polarized in the $x$-direction. The excitation pulse is resonant with the $1s^2\,\rightarrow\,1s3p$ transition, and fluorescence is measured from the $1s3p\,\rightarrow\,1s2s$ transition. 
	
\begin{figure}
	\centering
	\includegraphics[width=0.8\linewidth]{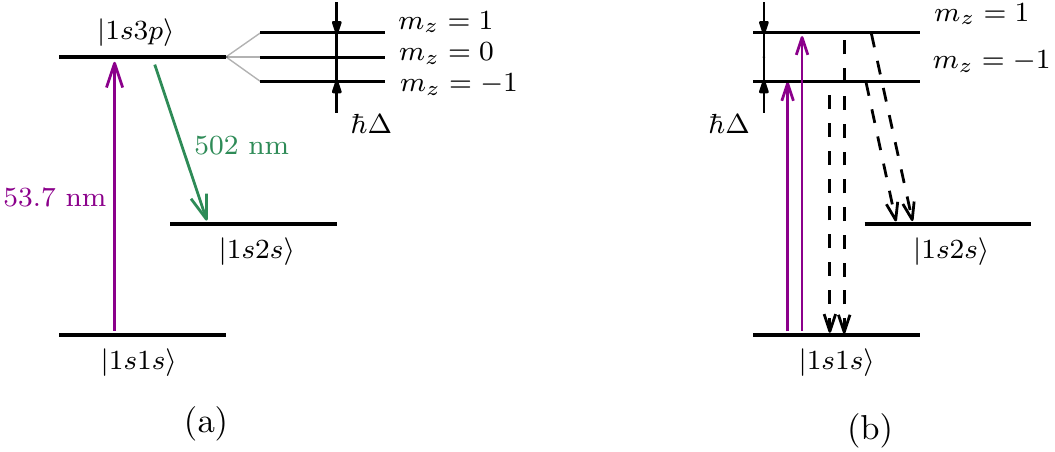}
	\caption{(a) Relevant excitation and fluorescence channels; (b) relevant transitions in the presence of a magnetic field.}
	\label{fig:levels}
\end{figure}

Due to the magnetic field of the MBS, the upper state transforms into a Zeeman triplet state (see Fig.\,\ref{fig:levels}\,(a)). The typical magnetic field strength of the permanent magnet in a MBS is a few tesla~\cite{tsuboi1988magnetic}, hence a rough estimation of the energy gap gives $\hbar\Delta \approx 0.1$ meV, which is about two orders of magnitude less than the FEL bandwidth ($\sim 30$ meV). Thus the magnetically split states are coherently excited.
	
Transitions from the Zeeman triplet substates to the singlet state $1s2s$ have different polarizations. The radial contributions to the matrix elements of the dipole moment on the $1s3p\,\rightarrow\,1s2s$ transition are denoted by $d$ and assumed to be real. Hence, the full matrix elements which take the polarization effects into account are:
	\begin{eqnarray}
	\langle m_z = -1| \widehat{\mathbf{d}} |1s2s\rangle = \frac{d}{\sqrt{2}} \left( \mathbf{e}_x + i \mathbf{e}_y \right),\\
	\langle m_z = 0| \widehat{\mathbf{d}} |1s2s\rangle = d \cdot \mathbf{e}_z,\\
	\langle m_z = +1| \widehat{\mathbf{d}} |1s2s\rangle = \frac{d}{\sqrt{2}} \left(-\mathbf{e}_x + i \mathbf{e}_y \right).
	\end{eqnarray}
Since the FEL pulse has linear polarization in the $x$-direction, the excitation process results in the population of the state
	\begin{equation}\label{eq:4}
	|m_x  = 0 \rangle = \frac{|m_z = -1\rangle - |m_z = + 1\rangle}{\sqrt{2}},
	\end{equation}
which is the eigenstate of spin-1 Pauli matrix $\sigma_x$ with zero projection $m_x = 0$. Hence, the level scheme is reduced to a four-level system containing one ground state, two intermediate excited states, and one final state (Fig.\,\ref{fig:levels}\,(b)) with the initial state described by Eq.\,(\ref{eq:4}). 
	
Since only single photon emission processes take place, the evolution of such a system can be solved analytically using the Weisskopf-Wigner approximation. For the case of ensemble of atoms, the single-atom contributions can be incoherently summed up. The resulting fluorescence intensity to the $1s2s$ state for the energy level scheme depicted in Fig.\,\ref{fig:levels}\,(b) is given by:
	\begin{equation}
	\label{I-vs-t_1}
	I(t) \sim \int d \Delta \cdot p(\Delta) \cdot e^{-\Gamma t} \cdot (1 + \cos \Delta \cdot t), 
	\end{equation}
where $t$ is the retardation time, $\Gamma$ is the total spontaneous decay rate of the upper states into the final states, and $p(\Delta)$ is the distribution function, which accounts for the inhomogeneity of the magnetic field strength within the interaction volume.

%For the case of ensemble of atoms, we assume 

%Since the total fluorescence intensity is proportional to the number of radiating atoms, the results from the single atom model are also applicable to a system of $N$ independent atoms, like our experiment.
	
\begin{figure}
		\centering 
		\includegraphics[width=0.65\linewidth]{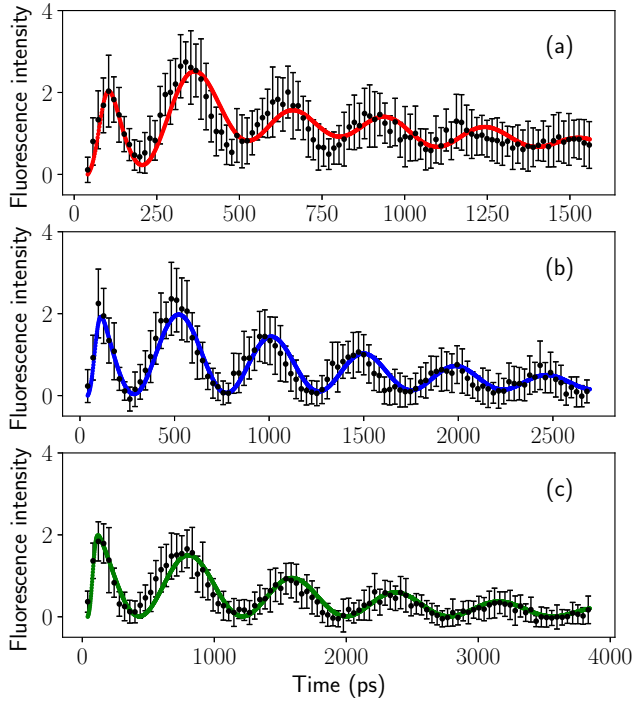}
		%\caption{Fitted fluorescence curves for different positions of the permanent magnet; (a) 6.5 mm from the interaction region with parameters $\Delta_0 =21.54 \cdot 10^{-3}$, $\sigma / N = 0.52  $; (b) 11.5 mm from the interaction region with parameters $\Delta_0 =  13.30 \cdot 10^{-3}$, $\sigma / N = 1.05$; (c) 16.5 mm from the interaction region with parameters $\Delta_0 =  8.12 \cdot 10^{-3}$, $\sigma / N = 1.78 $. The value of $\Delta_0$ is given in inverse picoseconds.}
		\caption{Fitted fluorescence curves for different positions of the permanent magnet; (a) 6.5 mm from the interaction region with parameters $\langle \Delta \rangle=18.6 \cdot 10^{-3}$, $\sigma_{\Delta} / \langle \Delta \rangle = 0.13  $; (b) 11.5 mm from the interaction region with parameters $\langle \Delta \rangle =  12.8 \cdot 10^{-3}$, $\sigma_{\Delta} / \langle \Delta \rangle = 0.03$; (c) 16.5 mm from the interaction region with parameters $\langle \Delta \rangle =  8.01 \cdot 10^{-3}$, $\sigma_{\Delta} / \langle \Delta \rangle = 0.012 $. The value of $\Delta$ is given in inverse picoseconds, $\sigma_{\Delta}=\sqrt{\langle \Delta^2 \rangle-\langle \Delta \rangle^2}$.}
		\label{fig:4}
\end{figure} 

%As one can see from the data, the period of the oscillations varies with the position of the magnet, which clearly indicates that the quantum beats originate from the Zeeman splitting of the excited state. The fluorescence intensity for three different positions of the magnet is shown in Fig.\,\ref{fig:2}\,(a-c). The significantly reduced intensity of the first peak in Fig.\,\ref{fig:2}\,(a) is due to the instrumental response function which has a width comparable to the period of the oscillations. As one can see from the data, the period of the oscillations varies with the position of the magnet, which clearly indicates that the quantum beats originate from the Zeeman splitting of the excited state. Since in our setup the polarization of the FEL pulses was perpendicular to the magnetic field generated by the magnetic bottle spectrometer (MBS), the atoms were excited to coherent superposition of two ($m=\pm 1$) excited states, which exhibits quantum beats. and the position of the magnet

Fig.\,\ref{fig:4} shows the temporal profile of the fluorescence intensity for three different positions for the permanent magnet; (a) 6.5 mm, (b) 11.5 mm, and (c) 16.5 mm from the interaction region. To model the experimental data presented in Fig.\,\ref{fig:4} we have used Eq.\,(\ref{I-vs-t_1}) and convoluted the result with a $20$ ps instrumental response function. For simplicity, within the interaction volume, we have assumed a Gaussian variation of magnetic field in space, and have recalculated the distribution function $p(\Delta)$ from it. The mean value and variance of the magnetic field distribution were used as fitting parameters, the spontaneous decay rate was taken fixed and equal to its theoretical value. 

As expected, for the positions of the magnet closer to the  interaction region, the beating period is shorter and the mean value of the magnetic field is larger. The inhomogeneity of the magnetic field strength within the interaction volume results in a smearing of the beating pattern. As one can see from the results of the fit, the magnetic field distribution is more inhomogeneous for closer positions of the magnet. This is as expected since the MBS is designed to confine electrons to helical orbits using a strongly divergent magnetic field.

\begin{figure}
	\centering 
	\includegraphics[width=0.9\linewidth]{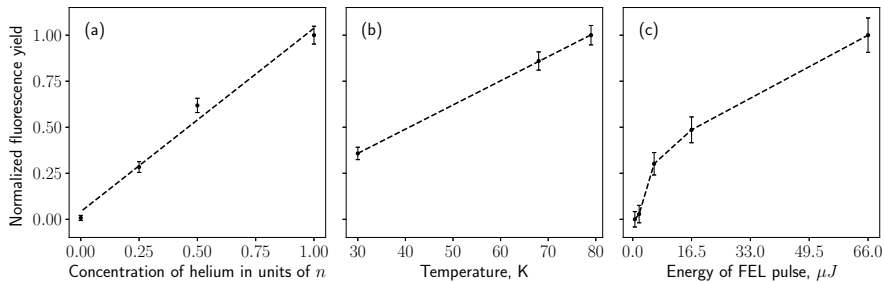}
	\caption{Fluorescence intensity as a function of different experimental parameters in normalized units; (a) varied atomic target density with fixed nozzle temperature $T = 79$ K and pulse energy $W = 66$ $\mu J$; (b) varied nozzle temperature for fixed density $n$ and pulse energy $W = 66$ $\mu J$; (c) varied excitation pulse energy with fixed temperature $T = 30$ K and concentration $n$. For these measurements, the permanent magnet was 6.5 mm from the interaction region.}
	\label{fig:3}
\end{figure} 
	
In our investigation, we additionally varied different experimental parameters, namely the target density and the nozzle temperature in the supersonic jet and the energy of the FEL excitation pulse. The total radiated energy may be estimated as the area under the fluorescence curve, which was measured for the different sets of parameters. Fig.\,\ref{fig:3}\,(a) shows that the total fluorescence intensity is linearly proportional to the atomic target density. A similar result is observed in Fig.\,\ref{fig:3}\,(b), which shows linear dependence on the nozzle temperature of the molecular beam source.  Fig.\,\ref{fig:3}\,(c) shows an increase of the total fluorescence intensity with pump pulse energy. 
%Since the pump pulse area is smaller than $\pi$ (here we understand the pump pulse area in the quantum-optical sense as the integral $\int dt\, \Omega(t)$, where $\Omega$ is Rabi-frequency), Rabi-like oscillatory behavior~\cite{Harries_2016_single-atom} is not expected, which is in agreement with our observation. 
The linear scaling with respect to atomic concentration shows that no collective effects -- such as superfluorescence -- have taken place. This is in contrast to previous experimental results showing superfluorescent behavior~\cite{Harries_2015} that were found for the same atomic transition, but without an external magnetic field. The primary reason is the target density from a supersonic gas jet used in this experiment is much less than that of previous experiments using a gas cell~\cite{Harries_2015, Harries_2016exp, Harries_2016th}.
%\textcolor{red}{ACL: What is What is meant by 'monotonous increase'? That term sounds strange to me. Also, it looks like the yield is saturating. Could that be the case?} 
%----------------------------------------------------------------

Overall, short wavelength, femtosecond pulses provided by XUV and X-ray FELs offer a novel means to initiate fluorescent quantum beats. The general requirement is that the pulses are i) short relative to the spontaneous decay time and beating period and ii) intense and broad enough to excite multiple excited states in order to produce the beat interference. Additionally, in contrast to multidimensional spectroscopy, a high degree of pump coherence is not necessary, as can be seen from fact that quantum beating was observed in synchrotron pulses \cite{hikosaka_zeeman_2020}. That said, FEL pulses offer a clear advantage of being much shorter and have higher peak intensities compared to synchrotron radiation. 

The quantum beat fluorescence approach with XUV/X-ray FEL pulses can be used to address excited states which have an energy difference $\Delta E$ such that beating with period $\hbar/\Delta E$ can be resolved with a streak camera. For a commercially available camera, this corresponds to about 100 fs, corresponding to $\Delta E$ below tens of meV. If the excited state decays more slowly than the beating period, the temporal oscillation of fluorescent radiation could be directly observed. A possible example would be valence double ionization of Xe. In this case, the branching ratio of this process  -- leading to excited Xe$^{2+}$ state --  is relatively large ($\sim\,30\%$~\cite{2017Khalal,2019Mercadier}). The lifetime of the valence-excited state levels is quite long, about 1 ns, hence if the excited state has a multiplet sub-structure with level spacing smaller than tens of meV, the fluorescence emitted on transition to the ground state of Xe$^{2+}$ could show quantum beating.

An extension to quantum beat fluorescence would be to increase the number of emitting atoms in order to drive the fluorescent radiation into the collective, superflourescent regime. As was previously shown in optical regime~\cite{vrehen1977quantum}, under the conditions of superfluorescence, the beating could originate not only from the splitting of the upper levels ($V$-level systems), but also from the splitting of the lower levels ($\Lambda$-level systems) where spontaneous fluorescence quantum beating is fundamentally impossible~\cite{scully1980quantum-beats}.

%The dependence on the pulse energy, shown in Fig.\,\ref{fig:3}\,(c), exhibits a fast rise which then transforms to the saturated square root dependence. %\textcolor{red}{Is there an explanation for this? I'll make a short discussion about an outlook of this technique tomorrow.} \textcolor{green}{We've made a quick estimation. For our experimental parameters area under the pulse seem to be much smaller that pi, which pretty much indicates that excitation efficiency is proportional to the pulse area. So the square root dependence we see may be the implication of area being proportional to $\sqrt{W}$ through amplitude.}

%In general, the short wavelength, femtosecond pulse durations provided by XUV and X-ray FELs give new opportunities to study quantum beating. The broadband nature of FEL pulses produced by self-amplified spontaneous emission presents the possibility to simultaneously excite multiple excited states with a single pulse. As a result, it is feasible to study quantum beating in new systems unavailable in the opitcal range. Furthermore, in the X-ray regime, excitation of core and inner valence shells are possible. In this case, one can additionally study quantum beating on different excitation schemes, where additional relaxation mechanisms such as Auger decay and autoionization are allowed.    	

\section{Conclusions}
In summary, we have used XUV radiation from a seeded FEL to resonantly excite the $1s3p$ state in atomic He. Using a femtosecond streak camera, we have observed time-resolved quantum beats in the fluorescence emission from the $1s3p\,\rightarrow\,1s2s$ transition due to the Zeemann splitting of the $1s3p$ state. To systematically understand the phenomenon, we have investigated the effect of the magnetic field, FEL pulse energy, and atomic source conditions on the fluorescence emission and quantum beating. The results demonstrate the feasibility of studying fluorescent quantum beating phenomena using ultrashort FEL pulses in the XUV and X-ray regime.

\section{Acknowledgements}
This work was funded by the Chemical Sciences, Geosciences, and Biosciences Division, Office of Basic Energy Sciences, Office of Science, US Department of Energy, Grant No. DESC0012376. R.W.F. acknowledges support from the Department of Energy, National Nuclear Security Administration Award DE-NA0003842 and Department of Energy, Office of Science, Office of Fusion Energy Sciences Award DESC0018298. R.M. and F.S. acknowledge support from the Deutsche Forschungsgemeinschaft under grant STI 125/19-2. The research leading to this result has been supported by the project CALIPSOplus under grant agreement 730872 from the EU Framework Programme for Research and Innovation HORIZON 2020.
\\

\bibliographystyle{unsrt} % Plain referencing style
\bibliography{sample} % Use the example bibliography file sample.bib

\end{document}